\documentclass[12pt,preprint]{aastex}
\usepackage{graphicx}

\begin{document}

\title{Galactic cluster winds in presence of a dark energy}

\author{Gennady S. BISNOVATYI-KOGAN\altaffilmark{1}}
\altaffiltext{1}{Space Research Institute (IKI)
\\ Profsoyuznaya 84/32, Moscow 117997, Russia
\\E-mail: gkogan@iki.rssi.ru}

\author{Marco MERAFINA\altaffilmark{2}}
\altaffiltext{2}{Department of Physics, University of Rome ``La Sapienza"
\\Piazzale Aldo Moro 2, I-00185 Rome, Italy
\\E-mail: marco.merafina@roma1.infn.it}

\begin{abstract}
We obtain a solution for the hydrodynamic outflow of the polytropic gas from the gravitating center, in presence of the uniform Dark Energy (DE). The antigravity of DE is enlightening the outflow and make the outflow possible at smaller initial temperature, at the same density. The main property of the wind in presence of DE is its unlimited acceleration after passing the critical point. In application of this solution to the winds from galaxy clusters we suggest that collision of the strongly accelerated wind with another galaxy cluster, or with another galactic cluster wind could lead to the formation of a highest energy cosmic rays.
\end{abstract}


\section{Introduction}

It was shown by Chernin (2001, 2008) that outer parts of galaxy clusters (GC) may be under strong influence of the dark energy (DE), discovered by observations of SN Ia at redshift $z\le 1$   (Riess et
al., 1998; Perlmutter et al., 1999), and in the spectrum of
fluctuations of the cosmic microwave background radiation (CMB), see
e.g. Spergel et al. (2003), Tegmark et al. (2004).  Equilibrium solutions for polytropic configurations in presence of DE have been obtained in papers of Balaguera-Antolinez et al. (2006, 2007), and Merafina et al. (2012).
The hot gas in the galactic clusters may flow outside due to high thermal pressure, and in the  outer parts of the cluster the presence of a dark energy (DE) facilitates the outflow.

Here we obtain a solution of hydrodynamic equations for the winds from galactic clusters in presence of DE. We generalize the solution for the outflows from the gravitating body, obtained for solar and stellar winds by Stanyukovich (1955) and Parker (1963), to the presence of DE. It implies significant changes in the structure of solutions describing galactic winds.

\section{Newtonian approximation in description galactic winds in presence of DE}

A transition to the Newtonian limit, where DE is described by the antigravity force in vacuum was done by Chernin (2008). In the Newtonian approximation, in presence of DE, we have the following hydrodynamic Euler equation for the spherically symmetric outflow in the gravitational field of matter and DE

\begin{equation}
\rho v\frac{dv}{dr}+ \frac{dP}{dr}=-\rho\left(\frac{Gm_m}{r^2}-\frac{\Lambda c^2 r}{3}\right)=-\rho\left(\frac{Gm_m}{r^2}-\frac{8\pi G \rho_\Lambda r}{3}\right).
 \label{eq1}
\end{equation}
Here $\rho$ and $P$ are a matter density and pressure, respectively, $m_m$ is the mass of the matter inside the radius $r$. We use here DE in the form of the Einstein cosmological constant $\Lambda$. Newtonian gravitational potentials produced by matter $\Phi_g$, and $\Phi_\Lambda$ by DE, satisfy the Poisson equations

\begin{equation}
\Delta\Phi_\Lambda =-8\pi G \rho_\Lambda, \quad  \Delta\Phi_g =4\pi G \rho, \quad \rho_\Lambda=\frac{\Lambda c^2}{8\pi G}.
 \label{eq2}
\end{equation}
We consider, for simplicity, the outflow in the field of a constant mass (like in stellar wind) $m_m=M$. The Eq. (\ref{eq1}) in this case is written as

\begin{equation}
\rho v\frac{dv}{dr}+ \frac{dP}{dr}=-\rho\left(\frac{GM}{r^2}-\frac{\Lambda c^2 r}{3}\right)=-\rho\left(\frac{GM}{r^2}-\frac{8\pi G \rho_\Lambda r}{3}\right).
 \label{eq3}
\end{equation}
The Eq. (\ref{eq1}) should be solved together with the continuity equation in the form

\begin{equation}
4\pi\rho v r^2=\dot M,
 \label{eq4}
\end{equation}
where $\dot M$ is the constant mass flux from the cluster. We consider polytropic equation of state, where pressure $P$, and sound speed $c_s$ are defined as

\begin{equation}
P=K\rho^{\gamma},\quad c_s^2=\gamma\frac{P}{\rho},\quad \rho=\left(\frac{c_s^2}{\gamma K}\right)^{\frac{1}{\gamma-1}},\quad P=\left(\frac{c_s^2}{\gamma }\right)^{\frac{\gamma}{\gamma-1}}K^{-\frac{1}{\gamma-1}}.
 \label{eq5}
\end{equation}
Introduce nondimensional variables as

\begin{equation}
\tilde v=\frac{v}{v_{*}}\quad \tilde c_s=\frac{c_s}{c_{*}}, \quad \tilde r=\frac{r}{r_*},\quad r_*=\frac{GM}{c_{*}^2},\quad v_*=c_*,
 \label{eq6}
\end{equation}
$$ \tilde\rho=\frac{\rho}{\rho_*}, \quad \tilde P=\frac{P}{P_*},\quad
\rho_*=\left(\frac{c_*^2}{\gamma K}\right)^{\frac{1}{\gamma-1}},\quad
P_*=\left(\frac{c_*^2}{\gamma }\right)^{\frac{\gamma}{\gamma-1}}K^{-\frac{1}{\gamma-1}}.
$$
In non-dimensional variables the equation  (\ref{eq3}) is written as

\begin{equation}
 \tilde v\frac{d\tilde v}{dr}+\frac{2}{\gamma-1}\tilde c_s\frac{d\tilde c_s}{d\tilde r}+\frac{1}{\tilde r^2}-\lambda \tilde r=0,\quad \lambda=\frac{\Lambda c^2 r_*^2}{3 c_*^2}.
 \label{eq7}
\end{equation}
The continuity equation (\ref{eq4}) in non-dimensional form is written as

\begin{equation}
\tilde\rho\,\tilde v\,\tilde r^2=\dot m,\quad
\tilde c_s^{\frac {2}{\gamma-1}} \tilde v \tilde r^2=\dot m, \quad
\dot m=\frac{\dot M}{\dot M_*},\quad
\dot M_*=4\pi\rho_* v_* r_*^2.
 \label{eq8}
\end{equation}
It follows from (\ref{eq5}),(\ref{eq6}),(\ref{eq8}), that

\begin{equation}
\frac{d\tilde\rho}{\tilde\rho}=\frac{2}{\gamma-1}\frac{d\tilde c_s}{\tilde c_s},\quad \frac{d\tilde\rho}{\tilde\rho}+\frac{d\tilde v}{\tilde v}+2\frac{d\tilde r}{\tilde r}=0.
 \label{eq9}
\end{equation}
Using (\ref{eq9}) we may write the equation of motion (\ref{eq7}) in the form

\begin{equation}
\frac{d\tilde v}{d\tilde r}=\frac{\tilde v}{\tilde r}\,\frac{2\tilde c_s^2-\frac{1}{\tilde r}+\lambda\tilde r^2}{\tilde v^2-\tilde c_s^2}.
 \label{eq10}
\end{equation}
The only physically relevant solutions are those which pass smoothly the sonic point $v=c_s$, being a singular point of the Eq. (10), with
\begin{equation}
\tilde v=\tilde c_s,\quad 2\tilde c_s^2-\frac{1}{\tilde r}+\lambda\tilde r^2=0
 \label{eq11}
\end{equation}
 where $\tilde r=\tilde r_c$,  $\tilde v=\tilde v_c$, $\tilde c_s=\tilde c_{sc}$. Choosing  $c_*=c_{sc}$, we obtain in the critical point

\begin{equation}
\tilde v_c=\tilde c_{sc}=1,\quad 2-\frac{1}{\tilde r_c}+\lambda\tilde r_c^2=0.
 \label{eq12}
\end{equation}
With this choice of the scaling paraneters, we have from (\ref{eq8})

\begin{equation}
\dot m=\tilde r_c^2.
 \label{eq12a}
\end{equation}
The relation (\ref{eq12}) determining the dependence $\tilde r_c(\lambda)$ in the solution for the galactic wind and accretion, in presence of DE, is presented in Fig.1.
The Eq.(\ref{eq7}) for the polytropic flow has a Bernoulli integral as

\begin{figure}[h]
\centerline{\includegraphics[scale=0.5]{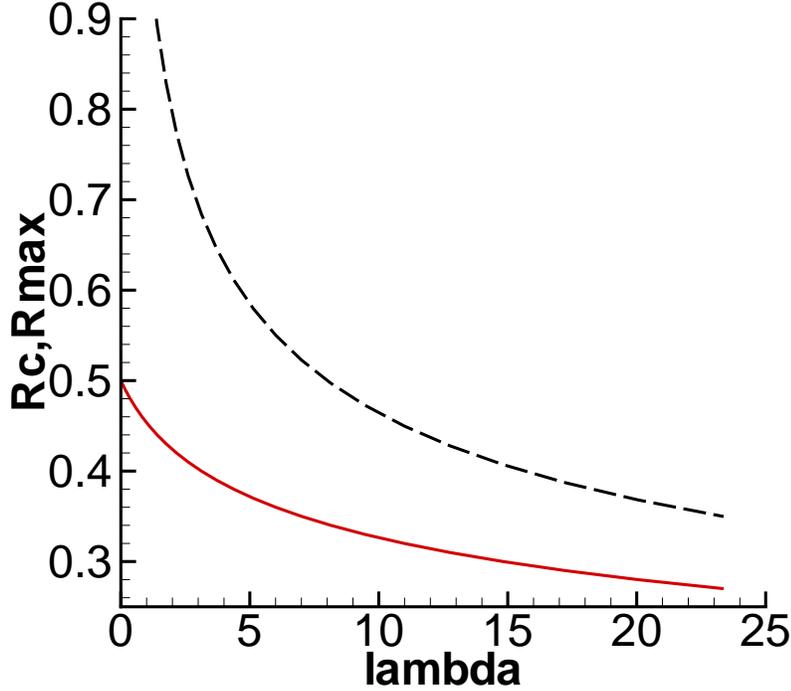} }
\figcaption{The functions $\tilde r_c(\lambda)$ (full curve), according to (\ref{eq12}), and $\tilde r_{max}(\lambda)$ (dashed curve) according to (\ref{eq20}). It is clear, that the critical radius of the flow $r_c$ is always inside the radius of
the extremum of the total gravitational potential $r_{max}$.
 \label{fig1}}
\end{figure}

\begin{equation}
\frac{\tilde v^2}{2}+\frac{\tilde c_s^2}{\gamma-1}-\frac{1}{\tilde r}-\frac{\lambda\tilde r^2}{2}=h , \quad \tilde c_s^2=\left(\frac{\dot m}{\tilde v \tilde r^2}\right)^{\gamma -1}=\left(\frac{\tilde r_c^2}{\tilde v \tilde r^2}\right)^{\gamma -1}.
 \label{eq13}
\end{equation}
 The dimensional Bernoulli integral $H=hc_{sc}^2$. The Bernoulli integral is determined through the parameters of the critical point, with account of (\ref{eq12}),  as

 \begin{equation}
h=\frac{\gamma+1}{2(\gamma-1)}-\frac{1}{\tilde r_c}-\frac{\lambda\tilde r_c^2}{2}=\frac{5-3\gamma}{2(\gamma-1)}-\frac{3}{2}\left(\frac{1}{\tilde r_c}-2\right).
 \label{eq14}
\end{equation}
The dependence $h(\lambda)$ for different polytropic powers $\gamma$ is given in Fig.2.

\begin{figure}[h]
\centerline{\includegraphics[scale=0.5]{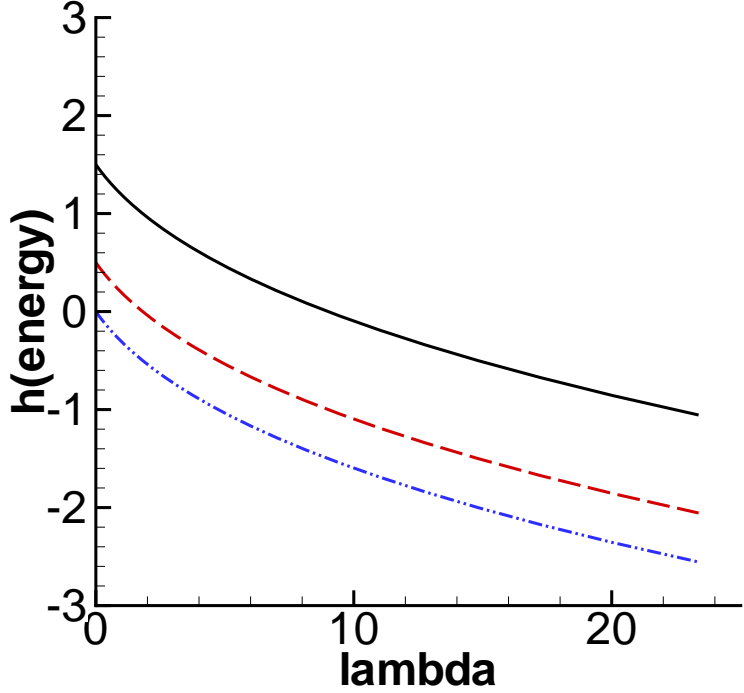} }
\figcaption{The function $h(\lambda)$ for $\gamma=\frac{4}{3}$ (full curve); $\gamma=\frac{3}{2}$ (dashed curve);
  $\gamma=\frac{5}{3}$ (dash-dot-dot curve), according to relations (\ref{eq12}),(\ref{eq14}).
 \label{fig2}}
\end{figure}
The stationary solution for the wind is determined by two integrals: constant mass flux $\dot M$, and energy (Bernoulli) integral $H$. In absence of DE we obtain the known relations

 \begin{equation}
\tilde r_c=\frac{1}{2},\quad h=\frac{5-3\gamma}{2(\gamma-1)}.
 \label{eq15}
\end{equation}
At small $\lambda$ we have from (\ref{eq12}),(\ref{eq14})

\begin{equation}
\tilde r_c=0.5-\frac{\lambda}{16},\quad h=\frac{5-3\gamma}{2(\gamma-1)}-\frac{3}{8}\lambda
\label{eq15b}
\end{equation}

At large $\lambda\rightarrow\infty$ it follows from  (\ref{eq12}) $\tilde r_c \rightarrow\tilde r_{c\infty}=\lambda^{-1/3}$. Making expansion in (\ref{eq12}) around  $\tilde r_{c\infty}$ in the form

$$\frac{1}{\tilde r_c}=\lambda^{1/3}+\varepsilon,$$
we obtain from (\ref{eq12}), (\ref{eq14})

 \begin{equation}
\varepsilon=\frac{2}{3},\quad \tilde r_c=\frac{1}{\lambda^{1/3}+\frac{2}{3}},\quad
 h=\frac{5-3\gamma}{2(\gamma-1)}-\frac{3}{2}\lambda^{1/3}+2=
 \frac{\gamma+1}{2(\gamma-1)}-\frac{3}{2}\lambda^{1/3}\quad {\rm at}
 \quad \lambda\rightarrow\infty.
 \label{eq15a}
\end{equation}
In the outflow from the physically relevant quasi-stationary object the antigravity from DE should be less the the gravitational force on the outer boundary, which we define at $r=r_*$. Therefore the value of $\Lambda$ is restricted by the relation (see e.g. Bisnovatyi-Kogan and Chernin, 2012)

 \begin{equation}
2\rho_\Lambda=\frac{\Lambda c^2}{4\pi G}<\bar\rho=\frac{4\pi M}{3r_*^3}
 \label{eq16}
\end{equation}
In non-dimensional variables this restriction, with account of (\ref{eq6}),(\ref{eq7}) is written as

  \begin{equation}
\lambda<\frac{16\pi^2}{9}=17.55=\lambda_{lim}.
 \label{eq17}
\end{equation}
 It is reasonable to consider only the values of $\lambda$ smaller than $\lambda_{lim}$.  It follows from (\ref{eq12}), that $\tilde r_c$ is monotonically decreasing with increasing $\lambda$. For $\lambda\,=\,\lambda_{lim}= 17.55$ we obtain $\tilde r_c=\tilde r_{c,lim}\approx 0.29$.
The effective gravitational potential $\tilde\Phi$ is formed by the gravity of the central body, and antigravity of DE

 \begin{equation}
\tilde\Phi=-\frac{1}{\tilde r}-\frac{\lambda \tilde r^2}{2}.
 \label{eq19}
\end{equation}
To overcome the gravity of the central body, the value of $h$ should exceed the maximum value of the gravitational potential, defined by the extremum of $\tilde\Phi$

 \begin{equation}
h\ge \tilde\Phi_{max}(\tilde r_{max})=-\frac{3}{2}\lambda^{1/3},\quad \tilde r_{max}=\lambda^{-1/3}.
 \label{eq20}
\end{equation}
The function $\tilde r_{max}(\lambda)$ is represented in Fig.1,  it is always $\tilde r_{max}>\tilde r_c$.
So, in presence of DE the outflow of the gas from the cluster  to the   infinity is possible even at the negative values of $h$. In absence of DE the non-negative value of $h$, and the outflow are possible only at $\gamma \le \frac{5}{3}$.

\section{Solutions of the galactic wind equation in presence of DE}

\subsection{Analytical solutions}

Analytical solutions exist only in absence of DE, $\Lambda=0$. In this case we have from (\ref{eq13})-(\ref{eq15})

\begin{equation}
\tilde r_c=\frac{1}{2},\quad h=\frac{\tilde v^2}{2}+\frac{\tilde c_s^2}{\gamma-1}-\frac{1}{\tilde r}=\frac{5-3\gamma}{2(\gamma-1)}.
 \label{eq23}
\end{equation}
For $\gamma=\frac{5}{3}$ the constant $h=0$ for the critical solution. At $h=0$ and $\gamma=\frac{5}{3}$ there is a whole family of solutions with arbitrary constant Mach number Ma=$\frac{v}{c_s}$ in all space.
It is more convenient to write this solution in dimensional variables, with Bernoulli integral

\begin{equation}
H=\frac{v^2}{2} + \frac{3c_s^2}{2}-\frac{GM}{r}=0.
 \label{eq24}
\end{equation}
There is an exact solution in the form

\begin{equation}
 v^2=\frac{2GM}{r}\frac{{\rm Ma}^2}{3+{\rm Ma^2}},\qquad c_s^2=\frac{2GM}{r}\frac{1}{3+{\rm Ma^2}}.
 \label{eq25}
\end{equation}
At Ma=1 this solution corresponds to the critical solutions for  $\gamma=\frac{5}{3}$.

Another analytic solution takes place at $\gamma=1.5$, with  the non-dimensional Bernoulli integral $h=0.5$.
Using from  (\ref{eq13}), for $\gamma=1.5$, $r_c=0.5$,

$$\tilde c_s^2=\left(\frac{\tilde r_c^2}{\tilde v \tilde r^2}\right)^{\gamma-1}=\sqrt{\frac{1}{4\tilde v \tilde r^2}},$$
we obtain
$$ h=\frac{\tilde v^2}{2}+\frac{1}{\tilde r\sqrt{\tilde v}}-\frac{1}{\tilde r}=\frac{1}{2}.$$
This nonlinear equation has two solutions

\begin{equation}
 {\bf 1.}\,\, \tilde v=1;\qquad {\bf 2.} \,\,\tilde r=\frac{2}{{\sqrt {\tilde v}}(1+\sqrt{{\tilde v}})(1+\tilde v)}.
 \label{eq26}
\end{equation}
There, the first solution corresponds to the wind, and the second one corresponds to the accretion case, with negative velocity;
 $\tilde v$ in (\ref{eq26}) represents the absolute value of this variable.

\subsection{Numerical solution}

To obtain a physically relevant critical solution of (\ref{eq10}), with $\tilde c_s^2$ from  (\ref{eq13}), we obtain expansion in the critical point with
$\tilde v^2=\tilde c_s^2=1$, in the form

\begin{equation}
\tilde v=1+\alpha(\tilde r-\tilde r_c),\quad \alpha_1=-\frac{2}{\tilde r_c}\frac{\gamma-1}{\gamma+1}+\frac{1}{\tilde r_c}\frac{2}{\gamma+1}
\sqrt{2+\frac{1}{4\tilde r_c}+\frac{\lambda \tilde r_c^2}{2}-\gamma\left(2-\frac{1}{4\tilde r_c}-\frac{\lambda \tilde r_c^2}{2}\right)},
 \label{eq27}
\end{equation}
$$ \qquad\qquad \alpha_2=-\frac{2}{\tilde r_c}\frac{\gamma-1}{\gamma+1}-\frac{1}{\tilde r_c}\frac{2}{\gamma+1}
\sqrt{2+\frac{1}{4\tilde r_c}+\frac{\lambda \tilde r_c^2}{2}-\gamma\left(2-\frac{1}{4\tilde r_c}-\frac{\lambda \tilde r_c^2}{2}\right)}.$$
Here $\alpha_1$ corresponds to the wind solution, and $\alpha_2$ is related to the case of accretion where $\tilde v$ define the absolute value.
At $\lambda=0$ we have a well known expansion with

$$\alpha_1=\frac{4}{\gamma+1}\left[\sqrt{\frac{5-3\gamma}{2}}-(\gamma-1)\right],$$
$$\alpha_2=-\frac{4}{\gamma+1}\left[\sqrt{\frac{5-3\gamma}{2}}+(\gamma-1)\right].$$
It follows from the expansion (\ref{eq27}), that physically relevant solutions exist only with positive value under the square root.
It give the restriction for the value of $\gamma$ as  a function of $\lambda$ in the form

$$\gamma\le \gamma_{max}=\frac{2+\frac{1}{4\tilde r_c}+\frac{\lambda \tilde r_c^2}{2}}{2-\frac{1}{4\tilde r_c}-\frac{\lambda \tilde r_c^2}{2}}.$$
At $\lambda=32$, $\tilde r_c=0.25$ the limiting value  $\gamma_{max}$ goes to $\infty$, so that at $\lambda \ge 32$ the wind solutions exist formally for all polytropic
powers $\gamma$. The dependence $\gamma_{max}(\lambda)$ is given in Fig.3.

\begin{figure}[h]
\centerline{\includegraphics[scale=0.5]{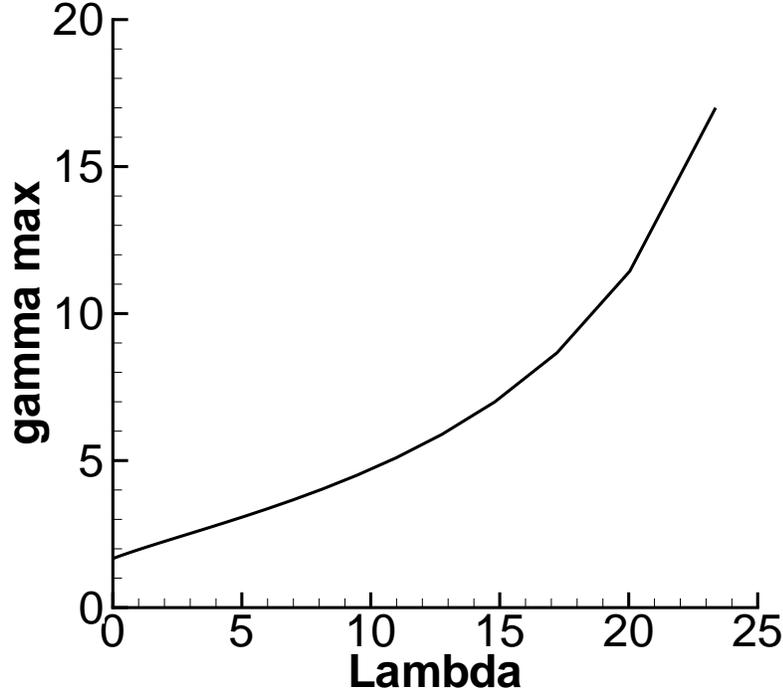} }
\figcaption{The functions $\gamma_{max}(\lambda)$. The smooth solution for the outflow from the gravitating center, in presence of DE, is possible only at $\gamma\le\gamma_{max}$.
 \label{fig3}}
\end{figure}
The critical solutions of the equation (\ref{eq10}), with account of (\ref{eq13}), are presented in Figs.4-6 for different values of $\gamma$ and $\lambda$. Both wind and accretion solutions are presented.

\begin{figure}[h]
\includegraphics[width=300pt]{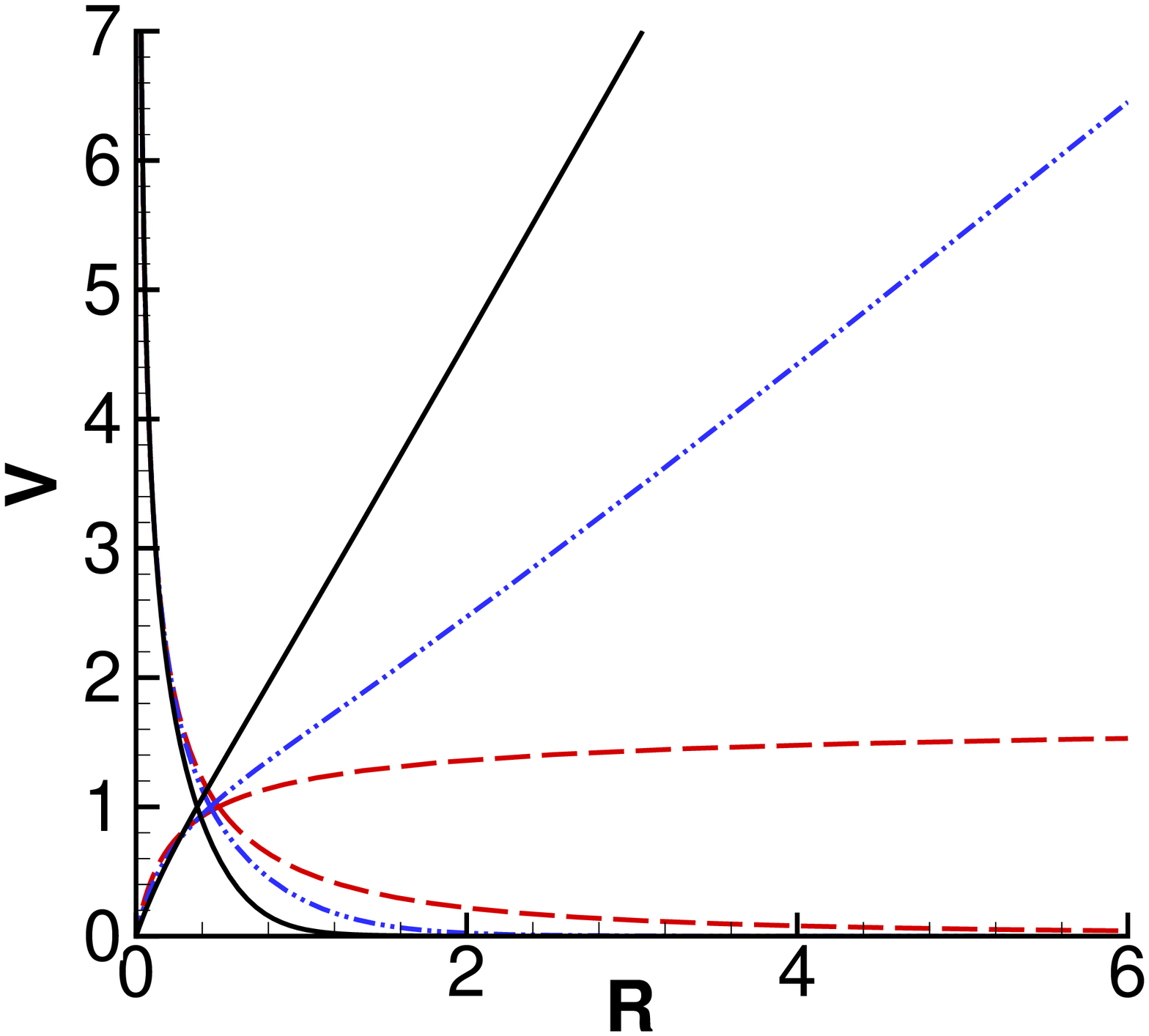}
\figcaption{The integral curves of the equations (\ref{eq10}), (\ref{eq13}), for $\gamma=4/3$ and $\lambda=0$, $r_c$=0.5 (dashed curves); $\lambda=1.10$, $r_c=0.45$ (dash-dot-dot curves); and $\lambda=5.13$, $r_c=0.37$ (full curves). Wind solutions correspond to curves with increasing velocity at large radius. The curves with decreasing velocities correspond to the accretion solution with negative $v$, so that its absolute value is presented.
 \label{fig4}}
\end{figure}

\begin{figure}[h]
\includegraphics[width=300pt]{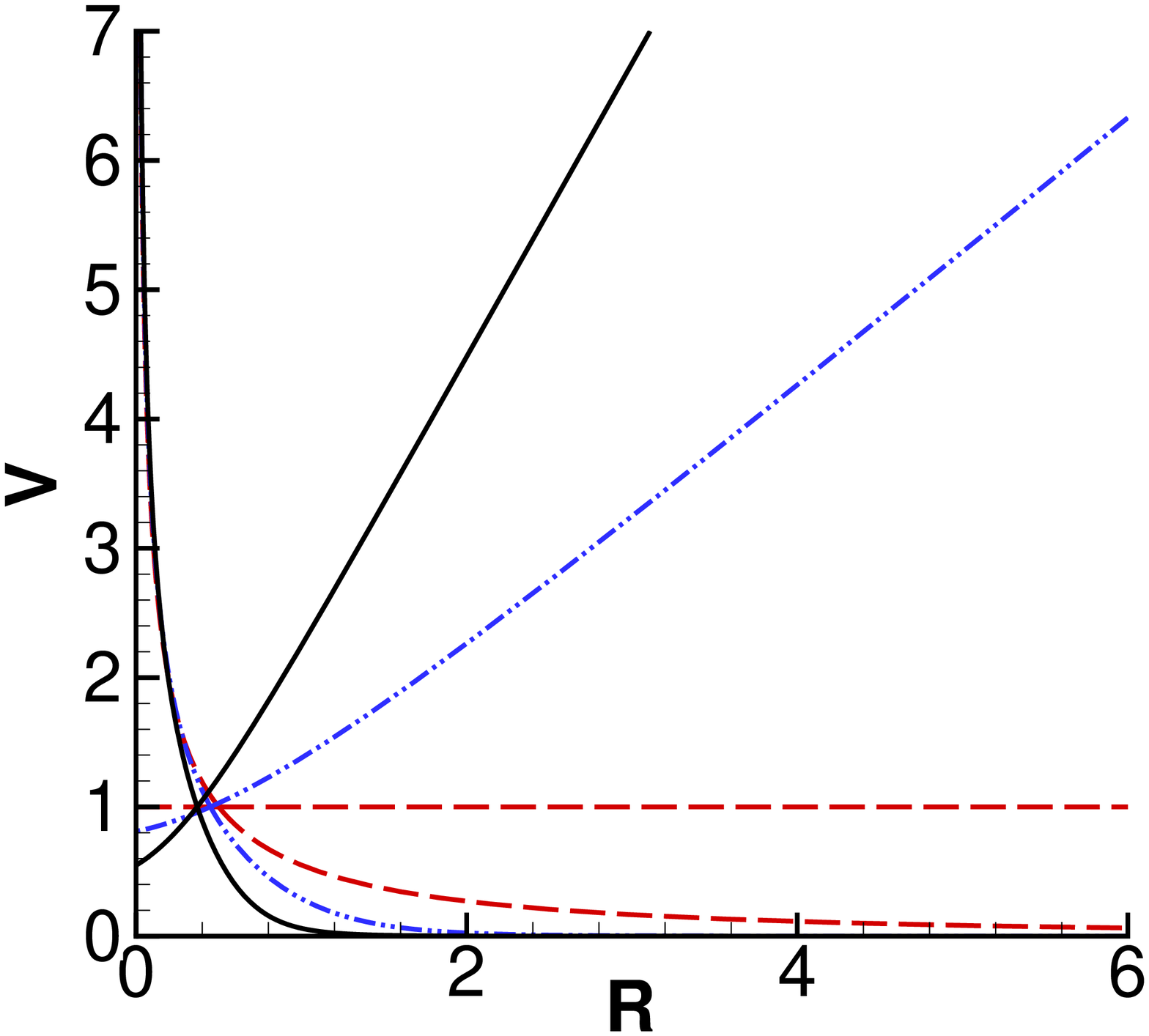}
\figcaption{The integral curves of the equations (\ref{eq10}), (\ref{eq13}), for $\gamma=3/2$ and $\lambda=0$, $r_c$=0.5 (dashed curves); $\lambda=1.10$, $r_c=0.45$ (dash-dot-dot curves); and $\lambda=5.13$, $r_c=0.37$ (full curves). Wind solutions correspond to curves with increasing (or constant) velocity at large radius. The curves with decreasing velocities correspond to the accretion solution with negative $v$, so that its absolute value is presented.
 \label{fig5}}
\end{figure}

\begin{figure}[h]
\includegraphics[width=300pt]{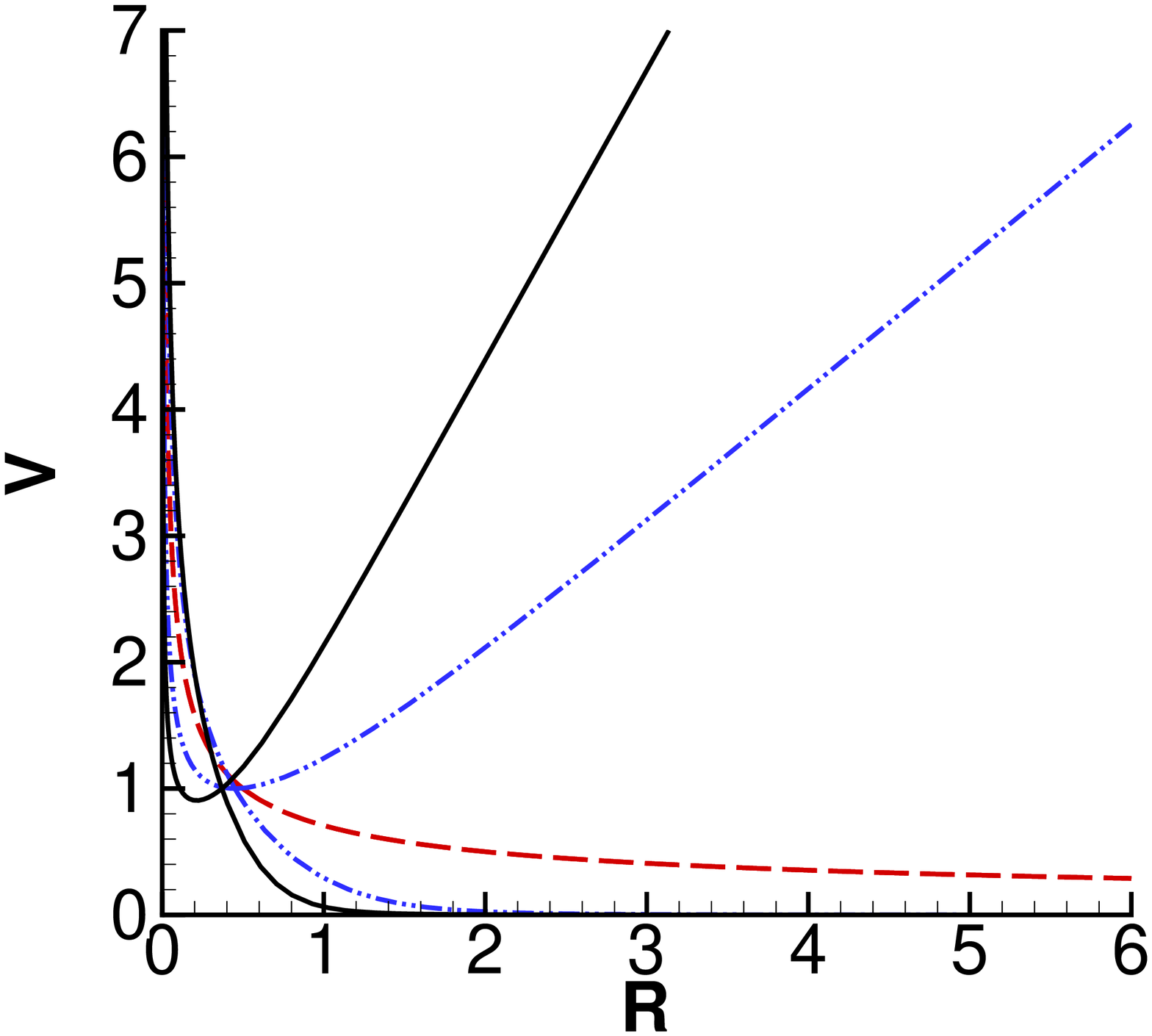}
\figcaption{The integral curves of the equations (\ref{eq10}), (\ref{eq13}), for $\gamma=5/3$ and $\lambda=0$, $r_c$=0.5 (dashed curve); $\lambda=1.10$, $r_c=0.45$ (dash-dot-dot curves); and $\lambda=5.13$, $r_c=0.37$ (full curves). For nonzero $\lambda$ wind solutions correspond to curves with increasing velocity at large radius. The curves with decreasing velocities correspond to the accretion solution with negative $v$, so that its absolute values are presented. At $\lambda=0$ both wind and accretion solutions are presented by the same curve, which corresponds to the wind for positive $v$, and to the accretion for negative $v$.
 \label{fig6}}
\end{figure}

\section{Discussion}

It is clear that the presence of DE tends to help the outflow of the hot gas from the gravitating object, as well as to the escape of rapidly moving galaxies (Chernin et al, 2013). Here we have obtained the solution for outflow in presence of DE, which generalize the well-known solution for the polytropic solar (stellar) wind. 
Presently the DE density exceed the density of the dark matter, and, even more, the density of the barionic matter. The clusters which outer radius is approaching the zero gravity radius, may not only loose  galaxies, which join the process of Hubble expansion, but also may loose the hot gas from the outer parts of the cluster.
Let us consider outer parts of the Coma cluster at radius $R_C=15$ Mpc, with the mass inside $M_C=5\cdot 10^{15}\,M_\odot$, from Chernin et al. (2013). For the present value of $\rho_\Lambda=0.71\cdot 10^{-29}$ g/cm$^3$, supposing that $R_C=r_*$ is the critical radius of the wind, we obtain from (2),(7), the nondimensional constant $\lambda$ as

\begin{equation}
\lambda=\frac{\Lambda c^2 r_*^2}{3 c_*^2}=\frac{8\pi}{3}\frac{\rho_\Lambda r_*^3}{M}\approx 0.59,\quad c_{*}=\sqrt{\frac{G\,M_C}{R_C}}\approx 1200\,\,{\rm km/c}.
 \label{eq28}
\end{equation}
It corresponds to the temperature about $T\approx 6\cdot 10^7$ K, $kT\approx 5$ keV. Observations of the hot gas distribution in the Coma cluster (Watanabe et al., 1999) on ASCA satellite have shown a presence of hot region with $kT=11-14$ keV, and more extended cool region with $kT=5\pm 1$ keV, what is in good accordance with our choice of parameters.

\begin{figure}[h]
\includegraphics[width=300pt]{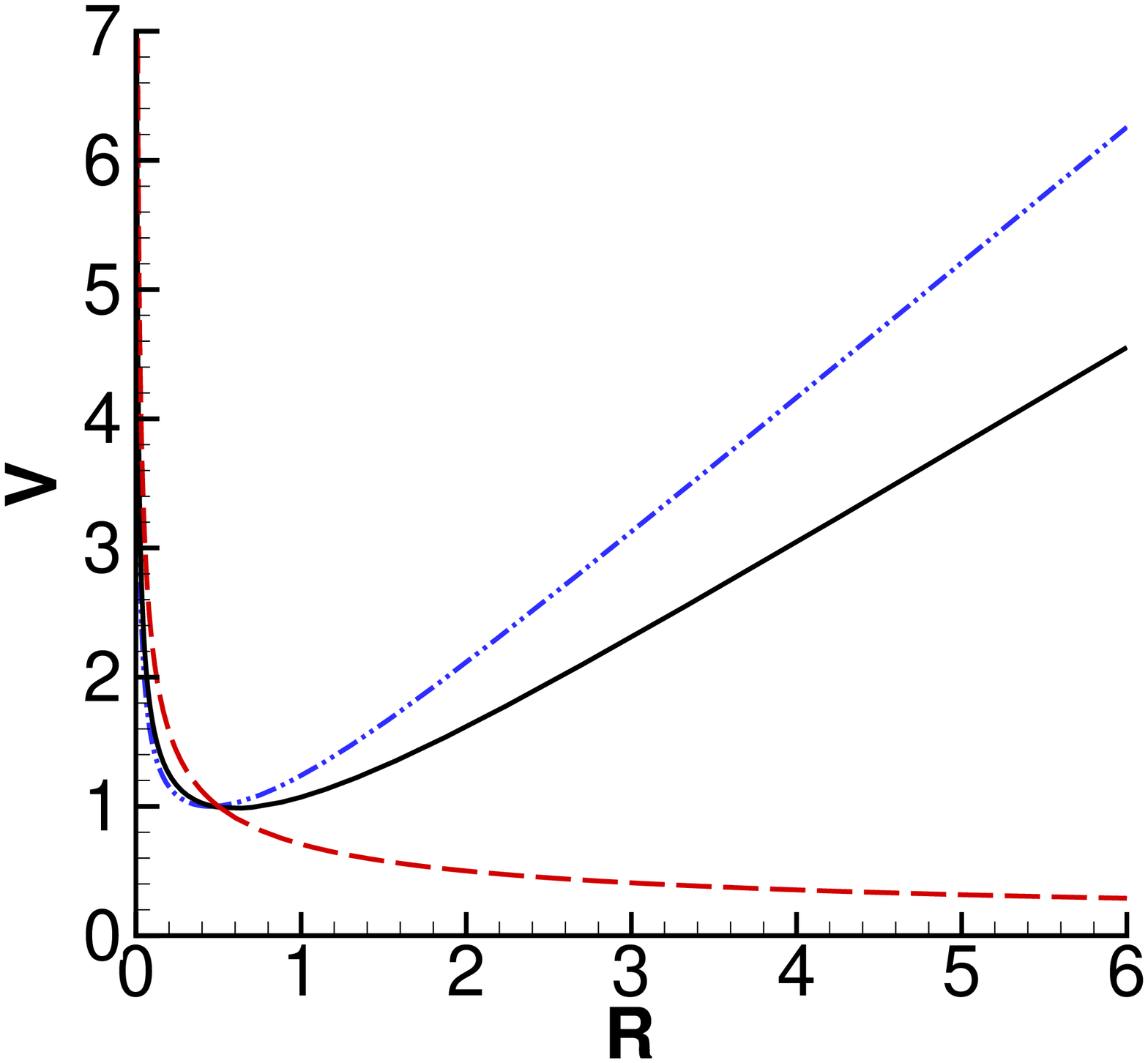}
\figcaption{The integral curves of the equations (\ref{eq10}), (\ref{eq13}) for the wind solution, at $\gamma=5/3$ and $\lambda=0$, $r_c$=0.5 (dashed curve); $\lambda=1.10$, $r_c=0.45$ (dash-dot-dot curves); and $\lambda=0.58$, $r_c=0.47$ (full curves). 
 \label{fig7}}
\end{figure}
Wind solutions for $\lambda$=0; 0.58; 1.1 are presented in Fig.7. The solution with $\lambda$=0.58 is the closest to the description of the outflow from Coma cluster. The density of the gas in the vicinity of $r=r_c$ is very small, so the flow may be considered as adiabatic (polytropic) with the power $\gamma$=5/3. Without DE such gas flow is inefficient, its velocity is decreasing $\sim 1/\sqrt{r}$, see Eq. (25). In presence of DE the wind velocity is increasing ~2 times at the distance of $\sim 5r_c \sim 75$ Mpc from Coma. 

After quitting the cluster the gas is moving with acceleration, acting as a snowplough for the intergalactic gas. The shell of matter, forming in such a way, may reach a  high velocity, exceeding considerably the speed of galaxies in cluster. If the shell meets another cluster, or another shell moving towards, the collision of such flows may induce a particle acceleration. Due to high speed, large sizes, and low density such collisions may create cosmic rays of the highest possible energy (EHECR). We may expect the largest effect when two clusters move to each other. The influence of DE is decreasing with with a red shift, therefore the acceleration of EHECR in this model should take place in the periphery, or between, the closest rich galaxy clusters.


\end{document}